\documentclass[%
 aip,
 jmp,%
 amsmath,amssymb,
 reprint,%
]{revtex4-1}

\usepackage{graphicx}
\usepackage{dcolumn}
\usepackage{bm}
\usepackage{sidecap}
\usepackage{ifthen}
\usepackage{float}

\begin{document}

\preprint{AIP/123-QED}

\title{Second Harmonic Generation in Photonic Crystal Cavities in (111)-Oriented GaAs}

\author{Sonia Buckley}
\email{bucklesm@stanford.edu.}
 \affiliation{E. L. Ginzton Laboratory, Stanford University, Stanford, CA 94305, U.S.A.}

\author{Marina Radulaski}
\affiliation{E. L. Ginzton Laboratory, Stanford University, Stanford, CA 94305, U.S.A.}

\author{Klaus Biermann}
\affiliation{Paul-Drude-Institut f\"{u}r Festk\"{o}rperelektronik,
Hausvogteiplatz 5-7 D-10117, Berlin, Germany}

\author{Jelena Vu\v{c}kovi\'{c}}
\affiliation{E. L. Ginzton Laboratory, Stanford University, Stanford, CA 94305, U.S.A.}

\begin{abstract}
We demonstrate second harmonic generation at telecommunications
wavelengths in photonic crystal cavities in (111)-oriented GaAs. We
fabricate 30 photonic crystal structures in both (111)- and (100)-
oriented GaAs and observe an increase in generated second harmonic power
in the (111) orientation, with the mean power increased by a factor of 3,
although there is a large scatter in the measured values.  We discuss
possible reasons for this increase, in particular the reduced two photon
absorption for transverse electric modes in (111) orientation, as well as
a potential increase due to improved mode overlap.
\end{abstract}

\maketitle

\begin{figure}
\includegraphics[width=8.5cm]{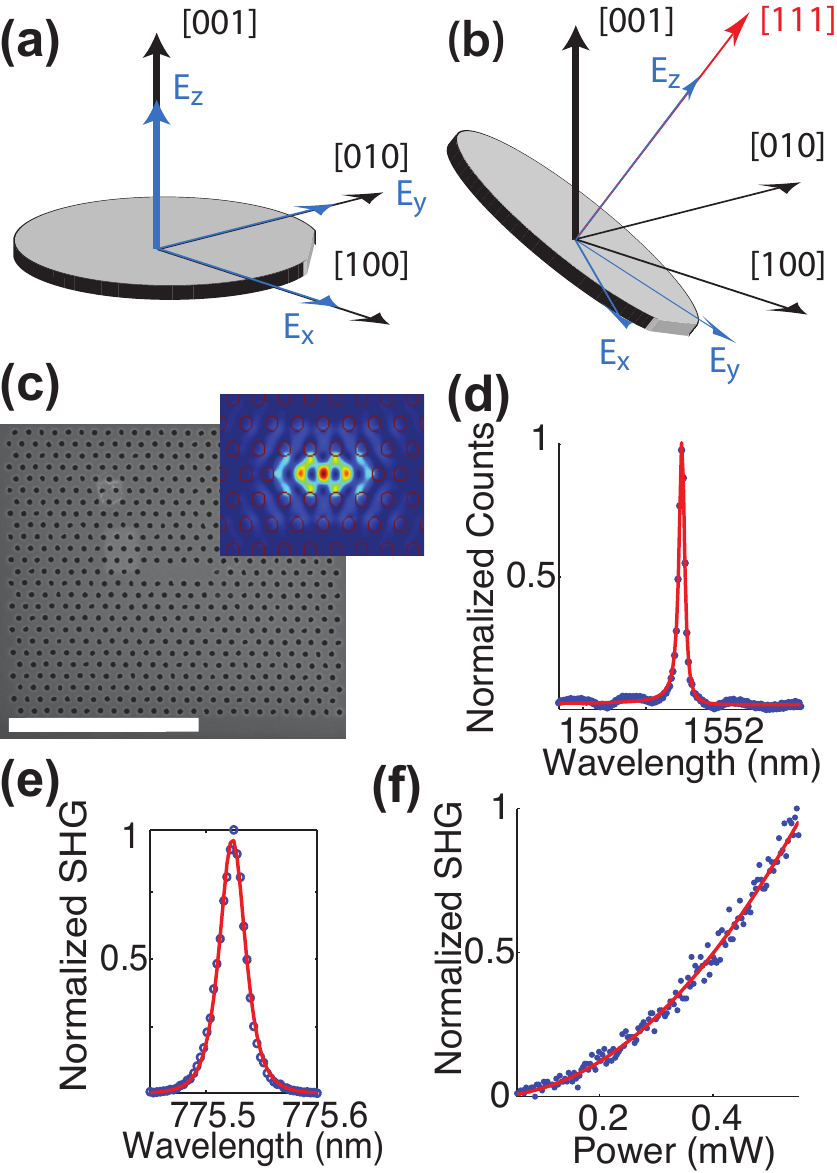}
\caption{(a) (001) and (b) (111) oriented III-V semiconductor wafer, showing electric field co-ordinates $E_x$, $E_y$ and $E_z$ aligned with the crystal axes.
For (111) orientation, each field component has components along all three of the main crystal axes.
(c) SEM of the three hole defect (L3) photonic crystal cavity fabricated in (111) GaAs. Simulated magnitude of the electric field of the fundamental mode is shown in the inset.  Scalebar is 5 $\mu$m.
(d) Cross polarized reflectivity spectrum of the cavity in (c) with resonance at 1551 nm. (e) Second harmonic spectrum of the cavity in (c) at 600 $\mu$W input power.
(f) Second harmonic power versus pump power for low pump powers for the same cavity as shown (c)-(e); red line shows quadratic fit.
 \label{fig:SEM}}
\end{figure}

Low mode-volume, high quality (Q) factor optical microcavities are promising
for nonlinear optical devices, as they have the potential to reach similar
conversion efficiencies to traditional optical cavities, but in
significantly more compact devices. In particular, if the optical mode
volume becomes small compared to the material nonlinear coherence length,
the phase matching condition is replaced by the requirement of large mode
overlap between the relevant optical modes \cite{rodriguez_chi2_2007}. This
is particularly useful in the case of III-V semiconductors such as GaAs, GaP
and InP, as these materials have a very high second order nonlinearity, but
do not exhibit birefringence.  This makes phasematching of bulk III-V
semiconductors challenging, and typically quasi-phasematching or additional
birefringent materials are employed. Quasi-phasematching these materials
poses additional challenges as, unlike frequently used LiNbO$_3$, III-Vs are
not ferroelectric and require wafer bonding or epitaxial growth for periodic
poling \cite{eyres_all-epitaxial_2001}. Conversely, integration of III-V
materials with optical microcavities is relatively easy due to the ease of
fabrication using semiconductor processing.  III-V semiconductors also allow
integration of active gain media such as quantum dots or quantum wells
\cite{rivoire_fast_2011,buckley_quasiresonant_2012,ota_nanocavity-based_2013},
as well as potential on-chip integration with detectors, switches and
modulators.

Experimentally, high efficiency, low power $\chi^{(2)}$ nonlinear processes
in resonant microcavities have been demonstrated, in particular second
harmonic generation in microdisks \cite{kuo_second-harmonic_2012} and
microrings \cite{levy_harmonic_2011}, as well as second harmonic generation
and sum frequency generation in photonic crystal cavities
\cite{mccutcheon_experimental_2007,
rivoire_second_2009,diziain_second_2013}. Millimeter sized lithium niobate
microdisks have also been used for high efficiency second harmonic
generation and ultra-low threshold optical parametric oscillators (OPOs)
\cite{ilchenko_low-threshold_2003,furst_naturally_2010,fortsch_versatile_2013}.
The efficiency of the conversion process is proportional to the quality
factors and spatial overlap integral of the three modes involved
 \cite{liscidini_highly_2004,rodriguez_chi2_2007}. In the
case of second harmonic generation in photonic crystal cavities, the
experimentally achieved efficiency of the processes has been limited by the
difficulty of engineering multiple high quality factor modes with a high
degree of overlap \cite{mccutcheon_experimental_2007,
rivoire_second_2009,diziain_second_2013}. This difficulty arises because the
bandgap of photonic crystals does not span a sufficiently large frequency
range for $\chi^{(2)}$ processes, and there are no significant higher order
bandgaps. This means that only one of the modes of the photonic crystal
cavity in the process is well defined and has high Q. For example, in
experimentally demonstrated second harmonic generation in three hole defect
(L3) photonic crystal cavities, the pump is coupled to the high Q
fundamental cavity mode. Previous work indicates that second harmonic
generation in photonic crystal cavities is primarily due to the bulk
$\chi^{(2)}$ rather than a surface $\chi^{(2)}$ effect (even in materials
where the generated second harmonic is above the bandgap and absorbed,
)\cite{mccutcheon_experimental_2007, rivoire_second_2009}. This suggests
that the second harmonic couples to leaky air band modes perturbed by the
presence of the cavity
\cite{mccutcheon_experimental_2007,rivoire_second_2009,buckley_quasiresonant_2012,diziain_second_2013},
which have low Q factors, and potentially low overlap with the fundamental
mode. In order to improve the efficiency, there have been several proposals
for designing photonic crystal cavities with multiple high Q resonances and
large frequency separations
\cite{zhang_ultra-high-Q_2009,burgess_design_2009,thon_polychromatic_2010,
rivoire_multiply_2011, rivoire_multiply_2011-1}.

Additional complications in designing nonlinear cavities in III-V
semiconductors arise due to the symmetry of the $\chi^{(2)}$ tensor; the
only nonzero elements of the bulk $\chi^{(2)}_{ijk}$ have $i \neq j \neq k$,
where $i,j$ and $k$ are directions along the [100], [010] and [001] crystal
axes.  Furthermore, the mode overlap is proportional to $\Sigma_{ijk}\int dr
\chi_{ijk}^{(2)}E_{1i}E_{2j}E_{3k}$, where $E$ is a component of the
electric field and the numeric subscript describes a distinct mode in a
three wave mixing process. Spatially anti-symmetric products of the three
field components have zero overlap. The vertical symmetry of the photonic
crystal cavity forces all modes to be either primarily transverse electric
(TE) -like or transverse magnetic (TM)
-like\cite{joannopoulos_photonic_2011}. For a TE-like mode, $E_x$, $E_y$ and
$H_z$ obey even symmetry in the vertical direction about the center of the
slab while $E_z,H_x,H_y = 0$ at the central $xy$ plane and are
anti-symmetric about this plane (where we define $z$ as the direction normal
to the wafer, and $x$ and $y$ to be in the plane of the wafer). For TM-like
modes, $E_z$, $H_x$ and $H_y$ obey even symmetry in the vertical direction
about the center of the slab while $E_x,E_y,H_z = 0$ at the central $xy$
plane and are anti-symmetric about this plane. Examining the mode overlap
integral reveals that for the process of second harmonic generation, a
TE-like mode may only couple to a TM-like mode if the wafer is normal to the
[100], [010] or [001] (equivalent) directions, as in standard (001) oriented
wafers \cite{rivoire_second_2009}. This is illustrated in Fig. \ref{fig:SEM}
(a). The consequence of this is that both high Q TE-like and TM-like modes
must be engineered for second harmonic generation in suspended photonic
crystals on this substrate, which is challenging as nearly all planar
photonic crystals only have a bandgap for TE polarization (with the
exception of a few recent demonstrations
\cite{takayama_experimental_2005,mccutcheon_high-q_2011}, which require
either complex 2D patterns or thick 1D structures, thereby making
fabrication more challenging ). However, if the wafer is cut such that the
plane of the wafer is normal to a different crystallographic direction, the
$x$ and $y$ directions in the plane of the wafer may have components along
each of the [100], [010] and [001] directions, and in this case TE-TE
coupling between modes may be allowed. For example, in the case of (111)
oriented III-Vs, illustrated in Fig. \ref{fig:SEM} (b), such TE-TE mode
coupling is allowed. This opens up additional degrees of freedom for
engineering photonic crystal cavities for frequency conversion
\cite{rivoire_multiply_2011-1}, as all three modes in a three wave mixing
process may now have either TE-like or TM-like symmetry.

In order to achieve efficient nonlinear frequency conversion, it is also
desirable to choose a semiconductor with a transparency window that overlaps
well with the experimental frequencies. GaP has a large bandgap and has been
demonstrated as a high-efficiency nonlinear source of visible frequencies
\cite{rivoire_second_2009}, but is challenging to grow in (111) orientation.
On the other hand, GaAs has a strong nonlinearity, is easier to grow in
(111) crystal orientation, and is compatible with bright gain media such as
InGaAs quantum wells, and efficient quantum emitters, such as InAs quantum
dots \cite{rivoire_fast_2011,ota_nanocavity-based_2013}. However, GaAs has a
smaller bandgap than GaP, and suffers from a large two photon absorption at
telecommunications wavelengths, as well as linear absorption of visible
wavelengths.  Moreover, due to the tensor nature of the nonlinear
properties, the two photon absorption of TE modes in (111) GaAs is several
times smaller than in (001) GaAs \cite{dvorak_measurement_1994}.

In order to demonstrate the improvement of nonlinear frequency conversion in
nanophotonic structures by controlling the crystal orientation of the
underlying material, we fabricate L3 photonic crystal cavities in (001) and
(111) GaAs and compare second harmonic generation in both. Photonic crystal
cavities were fabricated in 165 nm thick (001) and (111) GaAs membranes
grown on a thick AlGaAs sacrificial layer (1 \textbf{$\mu$}m thick in the
(001) sample, 0.8 \textbf{$\mu$}m thick in the (111) sample) on n-type doped
substrates. The (111)B GaAs substrate was off-oriented by 2$^\circ$ towards
[2-1-1]. The cavities were fabricated using e-beam lithography to define the
pattern, followed by dry etching and HF wet etching to remove the
sacrificial layer. An SEM of a fabricated device is shown in Fig.
\ref{fig:SEM} (c).  The fabricated structures had lattice constant $a$ =
460-470 nm and hole radius r/a = 0.28. The electric field of the fundamental
mode of this cavity is shown inset in the figure, and was simulated by 3D
finite difference time domain (FDTD) method. Fabricated photonic crystal
cavities were all characterized experimentally at the fundamental (1st
harmonic) wavelength with a broadband LED light source using a cross
polarized reflectivity method \cite{altug_photonic_2005}. The spectrum for
the structure shown in Fig. \ref{fig:SEM} (c), measured by this method, is
shown in Fig. \ref{fig:SEM} (d) with a Q factor of 10,000.

\begin{figure}
\includegraphics[width=8.5 cm]{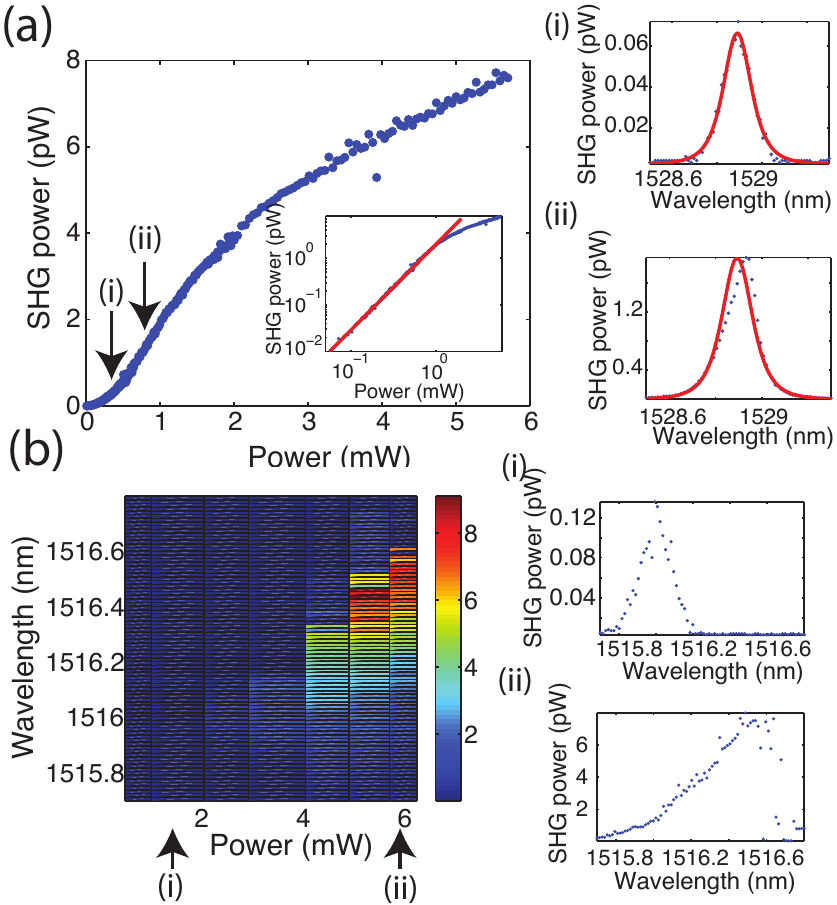}
{\caption{(a) Second harmonic generation in (111)-GaAs photonic crystal cavities versus input power.  Quadratic
dependence shown inset on loglog plot.  Power versus wavelength at different
power levels shown in (i) and (ii). (b) Second harmonic counts versus power
versus wavelength for a different structure.  (i) and (ii) show the power
versus wavelength for two different pump power levels.
\label{fig:cts_v_power}}}
\end{figure}

To perform second harmonic generation, a tunable CW telecommunications
wavelength (1465-1575 nm) laser is reflected from a 50-50 beamsplitter and
coupled to the high-Q fundamental cavity mode at normal incidence with a
high numerical aperture objective lens. The second harmonic emission is
collected through the same objective and passes through the beamsplitter to
be measured on a liquid nitrogen cooled Si CCD spectrometer
\cite{rivoire_second_2009}.  The tunable laser is scanned across the cavity
resonance, and the maximum counts measured on the spectrometer for each
laser wavelength are recorded. The value in counts/second was calibrated in
order to calculate actual powers. An example of such a spectrum for the
cavity in (111) GaAs measured shown in Fig. \ref{fig:SEM} (c) at 600 $\mu$W
pump power is shown in Fig. \ref{fig:SEM} (e), the red line is a fit to a
Lorentzian squared with Q factor of 10,000. The second harmonic power
dependence on input power can be recorded by varying the laser power at a
particular wavelength; the resulting curve for the same cavity is shown in
part (f), with a quadratic fit in red. Since the generated  second harmonic
is above the GaAs bandgap, the process efficiency is greatly reduced by the
two photon absorption of the pump and linear absorption of the second
harmonic; this is indicated by the fact that the measured power is
significantly lower than that measured from similar cavities in GaP, despite
the relatively higher $\chi^{(2)}$ nonlinearity of GaAs
\cite{shoji_absolute_1997}. Since at low powers the measured second
harmonic versus pump power is quadratic and the efficiency versus pump power
is approximately linear, in this power regime we can obtain the (constant)
conversion efficiency per Watt of input pump power for each structure. The
maximum low power conversion efficiency per Watt measured in a (111)-GaAs
structure was 0.002 \%/W (not accounting for in/out – coupling losses),
which is 450 times lower than the measured conversion efficiency per Watt in
GaP photonic crystal cavities.

Next, we compare the second harmonic spectra for similar L3 cavities
in (111) and (001) GaAs, in order to demonstrate the difference in the
second harmonic process in the two orientations for input powers leading to
collected second harmonic powers in the pW range.  We expect to observe a
difference in the measured second harmonic power for fixed pump power, due
to differences of the nonlinear optical interaction strengths and of the
two-photon absorption coefficients between the two samples, although this
experiment will not distinguish between the two effects.  Around 30
cavities were fabricated for both (111) and (001) GaAs. The input power was
kept at 600 $\mu$W, where the second harmonic generation (SHG) versus pump
curve was still quadratic for the case of (111) GaAs, as in Fig.
\ref{fig:SEM} (e), but the power was high enough that second harmonic signal
could be measured for most cavities. For (001) GaAs this may have already
been in the strong nonlinear absorption regime due to the higher two photon
absorption, however, below this power the second harmonic power was
too low to measure for a significant number of structures in the (001)
orientation, and therefore we chose to measure at 600 $\mu$W in order to
keep these structures in our sample.  We therefore do not quote low power
conversion efficiency for both but rather measure second harmonic power at
600 $\mu$W. For each photonic crystal cavity, the tunable laser was scanned
across the resonance. Second harmonic power was were plotted versus
wavelength, and the peak value was recorded. Table 1 summarizes the results,
which indicate higher efficiency SHG in the (111) orientation. A large
structure to structure variation was observed, and is possibly due to the
high sensitivity of air band mode frequencies to structure parameters, as
well as variations in in- and out-coupling efficiency. In order to try to
suppress this variation, we also fabricated perturbed L3 photonic crystal
cavities \cite{buckley_suppinfo_2013}. However, we still observed
significant scattering of values. Since this scattering was less severe in
the case of GaP cavities in previous studies \cite{rivoire_second_2009}
(e.g. allowing a noticeable dependence on in plane cavity rotation), we
suggest that the absorption contributes to the high sensitivity of measured
powers. The results for (001)-oriented GaAs are statistically enhanced by
the fact that there were several cavities in that orientation for which we
could not measure any SHG, perhaps because the efficiency was too low, and
these were not included in the results. Different rotations relative to the
crystal axes were also fabricated, however, the variation from structure to
structure was larger than any rotation variation that we could measure.
Therefore, we include structures of all rotations in the table.  The quality
factor of the structures measured was between 8,000 and 13,000, and no
significant dependence on Q factor was found in this range. We attribute
this to the fact that although the conversion efficiency increases with
increase in Q factor, the coupling efficiency decreases with increasing Q
factor, and in this range the two effects cancel. The increase in the SHG in
(111) cavities relative to (001) cavities may be a combination of the
smaller two-photon absorption in (111) GaAs and the possible better overlap
between the fundamental L3 cavity mode and air band modes, due to the
possibility of coupling to TE modes, as explained above. FDTD simulations of
the potential contributing air band modes \cite{buckley_quasiresonant_2012}
were used for calculations of the mode overlap, and indicate that this
overlap can be improved or reduced with a change in wafer orientation by up
to an order of magnitude, depending on the particular air band modes
involved. Determination of which of these modes is involved is challenging,
as these modes are low Q and therefore difficult to isolate via FDTD
simulation, and in general confirmation via farfield measurements and
rotation of the wafer are required.  In these simulations however, linear
and nonlinear absorption were not included, and therefore identification of
the precise modes involved was not possible. Another potential difference is
the difference in outcoupling between TE and TM modes.  One might naively
expect that TM modes would couple out more poorly than the TE modes.
However, for the low Q air band modes we simulated (see reference
\cite{buckley_quasiresonant_2012} for an example of such modes), we found
that between 20-70 \% of the light couples into the NA of the objective lens
for both TE modes and TM modes, with more variation depending on the precise
mode than between TE and TM. A precise characterization of the improvement
resulting from the chosen wafer orientation would require further
experiments at longer pump wavelengths to distinguish between improvement
from lower two photon absorption and improved overlap due to TE-TE mode
coupling.

\begin{table}
\caption{Comparison of second harmonic power in pW for L3 cavities in (001) and (111) GaAs for 600 $\mu$W input power.}
\begin{ruledtabular}
\begin{tabular}{ l  c  r }
\\
   & (001) GaAs (pW) & (111) GaAs (pW) \\ \\
  max & 1.66 & 7.3 \\                       \\
  mean & 0.47 & 1.55 \\                       \\
  median & 0.11 & 1.03 \\                       \\
  standard deviation & 0.55 & 1.6 \\              \\
  \end{tabular}
  \end{ruledtabular}
\end{table}

At higher input pump powers, the generated second harmonic will deviate from
a quadratic dependence due to absorption processes and resonance shifts.
This is shown in Fig. \ref{fig:cts_v_power} (a). The fundamental is absorbed
via two photon absorption, while the second harmonic is absorbed linearly.
This absorption leads to the generation of free carriers and free carrier
absorption. Additionally, free carrier absorption causes a change in
refractive index causing the cavity to blueshift
\cite{bennett_carrier-induced_1990}, as well as heating of the cavity,
causing the cavity to redshift (microsecond timescale). These transient
effects cause regenerative oscillations or bistability, particularly on the
red side of the resonance
\cite{gibbs_chapter_1985,van_optical_2002,xu_carrier-induced_2006}, and are
of interest for switching and optical signal processing applications
\cite{almeida_all-optical_2004,nozaki_sub-femtojoule_2010,fan_all-silicon_2012,
buckley_suppinfo_2013}. In addition, photo-oxidation of the cavity causes a
slow and permanent blueshift of the cavity resonance by up to several nm
\cite{lee_local_2009}. The input powers at which this deviation from
quadratic behavior occurs varied significantly from cavity to cavity. The
described resonance shifts can cause problems during measurement, as high Q
cavities shift off resonance suddenly. Fig. \ref{fig:cts_v_power} (a) parts
(i) and (ii) show the second harmonic spectrum at low and higher power; at
higher powers the spectrum can be seen to be slightly `tilted' from the
Lorentzian squared fit due to nonlinear absorption effects.  Fig.
\ref{fig:cts_v_power} (b) shows the second harmonic power versus wavelength
versus input power for another cavity with higher absorption.  As the power
is increased, the spectrum deviates more and more from a Lorentzian squared.
This is shown in parts (i) and (ii) where the spectra are shown at the low
and high power ends.   The high power spectrum is highly asymmetric and the
peak has redshifted; the second harmonic also becomes bistable at this side
of the spectrum (see supplement).

In summary, we have demonstrated second harmonic generation in (111) GaAs
with the pump at telecommunications wavelengths. We observed higher
conversion efficiencies for the (111) GaAs than in (001) GaAs, which is most
likely due to the combination of reduced two photon absorption for TE-like
modes relative to (001) GaAs, and better overlap between involved (TE-like)
modes. Future experiments at longer pump wavelengths could assert which
effects are
more responsible for this difference.\\
\\
This work was supported by the National Science Foundation (NSF Grant ECCS-
10 25811), a National Science Graduate Fellowship, and Stanford Graduate
Fellowships.  This work was performed in part at the Stanford
Nanofabrication Facility of NNIN supported by the National Science
Foundation under Grant No. ECS-9731293, and at the Stanford Nano Center. JV
also thanks the Alexander von Humboldt Foundation for support.

\section*{Supplementary Information}

\begin{figure}
\includegraphics[width=8.5cm]{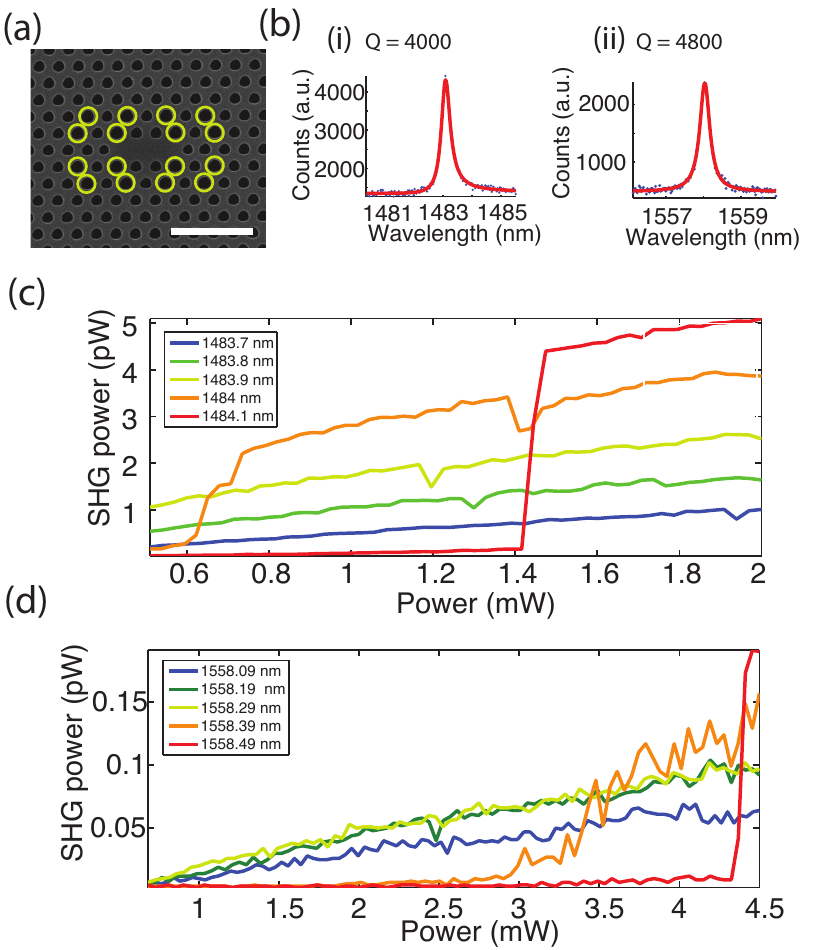}
\caption{(a) An SEM of a perturbed L3 cavity, enlarged holes are circled in yellow.  The scalebar is 2 $\mu$m.
(b) Cross-polarized reflectivity for a perturbed L3 cavity in (i) (111) GaAs and (ii) (001) GaAs.
(c)-(d) SHG power versus pump power for different wavelengths for (c) the (111) GaAs cavity characterized in (b) part (i),
 and (d) the (001) GaAs cavity characterized in (b) part (ii).
\label{fig:100_v_111}}
\end{figure}

To more easily characterize the bistable behavior of the cavities, we
fabricated perturbed L3 cavities in both (001) and (111) oriented wafers. An
SEM of a perturbed cavity is shown in Fig. \ref{fig:100_v_111} (a), the
perturbation is an increase in the radius of specific holes in the cavity
region indicated by the yellow circles. Incoupling is improved in these
cavities by folding back Fourier components outside the light cone to k = 0
using a perturbation of period $2a$, where $a$ is the lattice constant. This
degrades the Q factor; the cavities we measured had Q factors of
3,000-6,000. Fig. \ref{fig:100_v_111} (b) shows cross-polarized reflectivity
spectra for cavities in (i) (111) GaAs and (ii) (001) GaAs, with Q factors
of 4000 and 4800 respectively. The bistable behaviour was more easily
characterized in these cavities. Fig. \ref{fig:100_v_111} shows the SHG
power versus input power at different wavelengths for a typical perturbed
cavity in (c) (111) GaAs and (d) (001) GaAs. These are taken at a power
level where the dependence on power is no longer quadratic. The curves shown
on the blue side of the resonance are continuous, while the curves on the
red side of the resonance have a sudden jump in output power; this jump
corresponds to the onset of bistability in the structure. Oscillations will
occur at these wavelengths until photo-oxidation blueshifts the resonance
far enough and the power becomes unrecoverable. This happens as the cavity
heats up and the resonance redshifts and comes into resonance with the laser
wavelength. We note that the (001) cavity has lower counts than the (111)
cavity at all powers, which was typically the case.

\bibliographystyle{aipnum4-1}

\begin{thebibliography}{33}%
\makeatletter
\providecommand \@ifxundefined [1]{%
 \@ifx{#1\undefined}
}%
\providecommand \@ifnum [1]{%
 \ifnum #1\expandafter \@firstoftwo
 \else \expandafter \@secondoftwo
 \fi
}%
\providecommand \@ifx [1]{%
 \ifx #1\expandafter \@firstoftwo
 \else \expandafter \@secondoftwo
 \fi
}%
\providecommand \natexlab [1]{#1}%
\providecommand \enquote  [1]{``#1''}%
\providecommand \bibnamefont  [1]{#1}%
\providecommand \bibfnamefont [1]{#1}%
\providecommand \citenamefont [1]{#1}%
\providecommand \href@noop [0]{\@secondoftwo}%
\providecommand \href [0]{\begingroup \@sanitize@url \@href}%
\providecommand \@href[1]{\@@startlink{#1}\@@href}%
\providecommand \@@href[1]{\endgroup#1\@@endlink}%
\providecommand \@sanitize@url [0]{\catcode `\\12\catcode `\$12\catcode
  `\&12\catcode `\#12\catcode `\^12\catcode `\_12\catcode `\%12\relax}%
\providecommand \@@startlink[1]{}%
\providecommand \@@endlink[0]{}%
\providecommand \url  [0]{\begingroup\@sanitize@url \@url }%
\providecommand \@url [1]{\endgroup\@href {#1}{\urlprefix }}%
\providecommand \urlprefix  [0]{URL }%
\providecommand \Eprint [0]{\href }%
\providecommand \doibase [0]{http://dx.doi.org/}%
\providecommand \selectlanguage [0]{\@gobble}%
\providecommand \bibinfo  [0]{\@secondoftwo}%
\providecommand \bibfield  [0]{\@secondoftwo}%
\providecommand \translation [1]{[#1]}%
\providecommand \BibitemOpen [0]{}%
\providecommand \bibitemStop [0]{}%
\providecommand \bibitemNoStop [0]{.\EOS\space}%
\providecommand \EOS [0]{\spacefactor3000\relax}%
\providecommand \BibitemShut  [1]{\csname bibitem#1\endcsname}%
\let\auto@bib@innerbib\@empty
\bibitem [{\citenamefont {Rodriguez}\ \emph {et~al.}(2007)\citenamefont
  {Rodriguez}, \citenamefont {Solja\v{c}i\'{c}}, \citenamefont {Joannopoulos},\
  and\ \citenamefont {Johnson}}]{rodriguez_chi2_2007}%
  \BibitemOpen
  \bibfield  {author} {\bibinfo {author} {\bibfnamefont {A.}~\bibnamefont
  {Rodriguez}}, \bibinfo {author} {\bibfnamefont {M.}~\bibnamefont
  {Solja\v{c}i\'{c}}}, \bibinfo {author} {\bibfnamefont {J.~D.}\ \bibnamefont
  {Joannopoulos}}, \ and\ \bibinfo {author} {\bibfnamefont {S.~G.}\
  \bibnamefont {Johnson}},\ }\href {\doibase 10.1364/OE.15.007303} {\bibfield
  {journal} {\bibinfo  {journal} {Optics Express}\ }\textbf {\bibinfo {volume}
  {15}},\ \bibinfo {pages} {7303} (\bibinfo {year} {2007})}\BibitemShut
  {NoStop}%
\bibitem [{\citenamefont {Eyres}\ \emph {et~al.}(2001)\citenamefont {Eyres},
  \citenamefont {Tourreau}, \citenamefont {Pinguet}, \citenamefont {Ebert},
  \citenamefont {Harris}, \citenamefont {Fejer}, \citenamefont {Becouarn},
  \citenamefont {Gerard},\ and\ \citenamefont
  {Lallier}}]{eyres_all-epitaxial_2001}%
  \BibitemOpen
  \bibfield  {author} {\bibinfo {author} {\bibfnamefont {L.~A.}\ \bibnamefont
  {Eyres}}, \bibinfo {author} {\bibfnamefont {P.~J.}\ \bibnamefont {Tourreau}},
  \bibinfo {author} {\bibfnamefont {T.~J.}\ \bibnamefont {Pinguet}}, \bibinfo
  {author} {\bibfnamefont {C.~B.}\ \bibnamefont {Ebert}}, \bibinfo {author}
  {\bibfnamefont {J.~S.}\ \bibnamefont {Harris}}, \bibinfo {author}
  {\bibfnamefont {M.~M.}\ \bibnamefont {Fejer}}, \bibinfo {author}
  {\bibfnamefont {L.}~\bibnamefont {Becouarn}}, \bibinfo {author}
  {\bibfnamefont {B.}~\bibnamefont {Gerard}}, \ and\ \bibinfo {author}
  {\bibfnamefont {E.}~\bibnamefont {Lallier}},\ }\href {\doibase
  doi:10.1063/1.1389326} {\bibfield  {journal} {\bibinfo  {journal} {Applied
  Physics Letters}\ }\textbf {\bibinfo {volume} {79}},\ \bibinfo {pages} {904}
  (\bibinfo {year} {2001})}\BibitemShut {NoStop}%
\bibitem [{\citenamefont {Rivoire}\ \emph {et~al.}(2011)\citenamefont
  {Rivoire}, \citenamefont {Buckley}, \citenamefont {Majumdar}, \citenamefont
  {Kim}, \citenamefont {Petroff},\ and\ \citenamefont
  {Vu\v{c}kovi\'{c}}}]{rivoire_fast_2011}%
  \BibitemOpen
  \bibfield  {author} {\bibinfo {author} {\bibfnamefont {K.}~\bibnamefont
  {Rivoire}}, \bibinfo {author} {\bibfnamefont {S.}~\bibnamefont {Buckley}},
  \bibinfo {author} {\bibfnamefont {A.}~\bibnamefont {Majumdar}}, \bibinfo
  {author} {\bibfnamefont {H.}~\bibnamefont {Kim}}, \bibinfo {author}
  {\bibfnamefont {P.}~\bibnamefont {Petroff}}, \ and\ \bibinfo {author}
  {\bibfnamefont {J.}~\bibnamefont {Vu\v{c}kovi\'{c}}},\ }\href {\doibase
  doi:10.1063/1.3556644} {\bibfield  {journal} {\bibinfo  {journal} {Applied
  Physics Letters}\ }\textbf {\bibinfo {volume} {98}},\ \bibinfo {pages}
  {083105} (\bibinfo {year} {2011})}\BibitemShut {NoStop}%
\bibitem [{\citenamefont {Buckley}\ \emph {et~al.}(2012)\citenamefont
  {Buckley}, \citenamefont {Rivoire}, \citenamefont {Hatami},\ and\
  \citenamefont {Vu\v{c}kovi\'{c}}}]{buckley_quasiresonant_2012}%
  \BibitemOpen
  \bibfield  {author} {\bibinfo {author} {\bibfnamefont {S.}~\bibnamefont
  {Buckley}}, \bibinfo {author} {\bibfnamefont {K.}~\bibnamefont {Rivoire}},
  \bibinfo {author} {\bibfnamefont {F.}~\bibnamefont {Hatami}}, \ and\ \bibinfo
  {author} {\bibfnamefont {J.}~\bibnamefont {Vu\v{c}kovi\'{c}}},\ }\href
  {\doibase doi:10.1063/1.4761248} {\bibfield  {journal} {\bibinfo  {journal}
  {Applied Physics Letters}\ }\textbf {\bibinfo {volume} {101}},\ \bibinfo
  {pages} {161116} (\bibinfo {year} {2012})}\BibitemShut {NoStop}%
\bibitem [{\citenamefont {Ota}\ \emph {et~al.}(2013)\citenamefont {Ota},
  \citenamefont {Watanabe}, \citenamefont {Iwamoto},\ and\ \citenamefont
  {Arakawa}}]{ota_nanocavity-based_2013}%
  \BibitemOpen
  \bibfield  {author} {\bibinfo {author} {\bibfnamefont {Y.}~\bibnamefont
  {Ota}}, \bibinfo {author} {\bibfnamefont {K.}~\bibnamefont {Watanabe}},
  \bibinfo {author} {\bibfnamefont {S.}~\bibnamefont {Iwamoto}}, \ and\
  \bibinfo {author} {\bibfnamefont {Y.}~\bibnamefont {Arakawa}},\ }\href
  {\doibase 10.1364/OE.21.019778} {\bibfield  {journal} {\bibinfo  {journal}
  {Optics Express}\ }\textbf {\bibinfo {volume} {21}},\ \bibinfo {pages}
  {19778} (\bibinfo {year} {2013})}\BibitemShut {NoStop}%
\bibitem [{\citenamefont {Kuo}\ and\ \citenamefont
  {Solomon}(2012)}]{kuo_second-harmonic_2012}%
  \BibitemOpen
  \bibfield  {author} {\bibinfo {author} {\bibfnamefont {P.~S.}\ \bibnamefont
  {Kuo}}\ and\ \bibinfo {author} {\bibfnamefont {G.~S.}\ \bibnamefont
  {Solomon}},\ }\href {http://arxiv.org/abs/1210.1984} {\bibfield  {journal}
  {\bibinfo  {journal} {{arXiv:1210.1984}}\ } (\bibinfo {year}
  {2012})}\BibitemShut {NoStop}%
\bibitem [{\citenamefont {Levy}\ \emph {et~al.}(2011)\citenamefont {Levy},
  \citenamefont {Foster}, \citenamefont {Gaeta},\ and\ \citenamefont
  {Lipson}}]{levy_harmonic_2011}%
  \BibitemOpen
  \bibfield  {author} {\bibinfo {author} {\bibfnamefont {J.~S.}\ \bibnamefont
  {Levy}}, \bibinfo {author} {\bibfnamefont {M.~A.}\ \bibnamefont {Foster}},
  \bibinfo {author} {\bibfnamefont {A.~L.}\ \bibnamefont {Gaeta}}, \ and\
  \bibinfo {author} {\bibfnamefont {M.}~\bibnamefont {Lipson}},\ }\href
  {\doibase 10.1364/OE.19.011415} {\bibfield  {journal} {\bibinfo  {journal}
  {Optics Express}\ }\textbf {\bibinfo {volume} {19}},\ \bibinfo {pages}
  {11415} (\bibinfo {year} {2011})}\BibitemShut {NoStop}%
\bibitem [{\citenamefont {{McCutcheon}}\ \emph {et~al.}(2007)\citenamefont
  {{McCutcheon}}, \citenamefont {Young}, \citenamefont {Rieger}, \citenamefont
  {Dalacu}, \citenamefont {Fr\'{e}d\'{e}rick}, \citenamefont {Poole},\ and\
  \citenamefont {Williams}}]{mccutcheon_experimental_2007}%
  \BibitemOpen
  \bibfield  {author} {\bibinfo {author} {\bibfnamefont {M.~W.}\ \bibnamefont
  {{McCutcheon}}}, \bibinfo {author} {\bibfnamefont {J.~F.}\ \bibnamefont
  {Young}}, \bibinfo {author} {\bibfnamefont {G.~W.}\ \bibnamefont {Rieger}},
  \bibinfo {author} {\bibfnamefont {D.}~\bibnamefont {Dalacu}}, \bibinfo
  {author} {\bibfnamefont {S.}~\bibnamefont {Fr\'{e}d\'{e}rick}}, \bibinfo
  {author} {\bibfnamefont {P.~J.}\ \bibnamefont {Poole}}, \ and\ \bibinfo
  {author} {\bibfnamefont {R.~L.}\ \bibnamefont {Williams}},\ }\href {\doibase
  10.1103/PhysRevB.76.245104} {\bibfield  {journal} {\bibinfo  {journal}
  {Physical Review B}\ }\textbf {\bibinfo {volume} {76}},\ \bibinfo {pages}
  {245104} (\bibinfo {year} {2007})}\BibitemShut {NoStop}%
\bibitem [{\citenamefont {Rivoire}\ \emph {et~al.}(2009)\citenamefont
  {Rivoire}, \citenamefont {Lin}, \citenamefont {Hatami}, \citenamefont
  {Masselink},\ and\ \citenamefont {Vu\v{c}kovi\'{c}}}]{rivoire_second_2009}%
  \BibitemOpen
  \bibfield  {author} {\bibinfo {author} {\bibfnamefont {K.}~\bibnamefont
  {Rivoire}}, \bibinfo {author} {\bibfnamefont {Z.}~\bibnamefont {Lin}},
  \bibinfo {author} {\bibfnamefont {F.}~\bibnamefont {Hatami}}, \bibinfo
  {author} {\bibfnamefont {W.~T.}\ \bibnamefont {Masselink}}, \ and\ \bibinfo
  {author} {\bibfnamefont {J.}~\bibnamefont {Vu\v{c}kovi\'{c}}},\ }\href
  {\doibase 10.1364/OE.17.022609} {\bibfield  {journal} {\bibinfo  {journal}
  {Optics Express}\ }\textbf {\bibinfo {volume} {17}},\ \bibinfo {pages}
  {22609} (\bibinfo {year} {2009})}\BibitemShut {NoStop}%
\bibitem [{\citenamefont {Diziain}\ \emph {et~al.}(2013)\citenamefont
  {Diziain}, \citenamefont {Geiss}, \citenamefont {Zilk}, \citenamefont
  {Schrempel}, \citenamefont {Kley}, \citenamefont {T\"{u}nnermann},\ and\
  \citenamefont {Pertsch}}]{diziain_second_2013}%
  \BibitemOpen
  \bibfield  {author} {\bibinfo {author} {\bibfnamefont {S.}~\bibnamefont
  {Diziain}}, \bibinfo {author} {\bibfnamefont {R.}~\bibnamefont {Geiss}},
  \bibinfo {author} {\bibfnamefont {M.}~\bibnamefont {Zilk}}, \bibinfo {author}
  {\bibfnamefont {F.}~\bibnamefont {Schrempel}}, \bibinfo {author}
  {\bibfnamefont {E.-B.}\ \bibnamefont {Kley}}, \bibinfo {author}
  {\bibfnamefont {A.}~\bibnamefont {T\"{u}nnermann}}, \ and\ \bibinfo {author}
  {\bibfnamefont {T.}~\bibnamefont {Pertsch}},\ }\href {\doibase
  doi:10.1063/1.4817507} {\bibfield  {journal} {\bibinfo  {journal} {Applied
  Physics Letters}\ }\textbf {\bibinfo {volume} {103}},\ \bibinfo {pages}
  {051117} (\bibinfo {year} {2013})}\BibitemShut {NoStop}%
\bibitem [{\citenamefont {Ilchenko}\ \emph {et~al.}(2003)\citenamefont
  {Ilchenko}, \citenamefont {Matsko}, \citenamefont {Savchenkov},\ and\
  \citenamefont {Maleki}}]{ilchenko_low-threshold_2003}%
  \BibitemOpen
  \bibfield  {author} {\bibinfo {author} {\bibfnamefont {V.~S.}\ \bibnamefont
  {Ilchenko}}, \bibinfo {author} {\bibfnamefont {A.~B.}\ \bibnamefont
  {Matsko}}, \bibinfo {author} {\bibfnamefont {A.~A.}\ \bibnamefont
  {Savchenkov}}, \ and\ \bibinfo {author} {\bibfnamefont {L.}~\bibnamefont
  {Maleki}},\ }\href {\doibase 10.1364/JOSAB.20.001304} {\bibfield  {journal}
  {\bibinfo  {journal} {Journal of the Optical Society of America B}\ }\textbf
  {\bibinfo {volume} {20}},\ \bibinfo {pages} {1304} (\bibinfo {year}
  {2003})}\BibitemShut {NoStop}%
\bibitem [{\citenamefont {F\"{u}rst}\ \emph {et~al.}(2010)\citenamefont
  {F\"{u}rst}, \citenamefont {Strekalov}, \citenamefont {Elser}, \citenamefont
  {Lassen}, \citenamefont {Andersen}, \citenamefont {Marquardt},\ and\
  \citenamefont {Leuchs}}]{furst_naturally_2010}%
  \BibitemOpen
  \bibfield  {author} {\bibinfo {author} {\bibfnamefont {J.~U.}\ \bibnamefont
  {F\"{u}rst}}, \bibinfo {author} {\bibfnamefont {D.~V.}\ \bibnamefont
  {Strekalov}}, \bibinfo {author} {\bibfnamefont {D.}~\bibnamefont {Elser}},
  \bibinfo {author} {\bibfnamefont {M.}~\bibnamefont {Lassen}}, \bibinfo
  {author} {\bibfnamefont {U.~L.}\ \bibnamefont {Andersen}}, \bibinfo {author}
  {\bibfnamefont {C.}~\bibnamefont {Marquardt}}, \ and\ \bibinfo {author}
  {\bibfnamefont {G.}~\bibnamefont {Leuchs}},\ }\href {\doibase
  10.1103/PhysRevLett.104.153901} {\bibfield  {journal} {\bibinfo  {journal}
  {Physical Review Letters}\ }\textbf {\bibinfo {volume} {104}},\ \bibinfo
  {pages} {153901} (\bibinfo {year} {2010})}\BibitemShut {NoStop}%
\bibitem [{\citenamefont {F\"{o}rtsch}\ \emph {et~al.}(2013)\citenamefont
  {F\"{o}rtsch}, \citenamefont {F\"{u}rst}, \citenamefont {Wittmann},
  \citenamefont {Strekalov}, \citenamefont {Aiello}, \citenamefont {Chekhova},
  \citenamefont {Silberhorn}, \citenamefont {Leuchs},\ and\ \citenamefont
  {Marquardt}}]{fortsch_versatile_2013}%
  \BibitemOpen
  \bibfield  {author} {\bibinfo {author} {\bibfnamefont {M.}~\bibnamefont
  {F\"{o}rtsch}}, \bibinfo {author} {\bibfnamefont {J.~U.}\ \bibnamefont
  {F\"{u}rst}}, \bibinfo {author} {\bibfnamefont {C.}~\bibnamefont {Wittmann}},
  \bibinfo {author} {\bibfnamefont {D.}~\bibnamefont {Strekalov}}, \bibinfo
  {author} {\bibfnamefont {A.}~\bibnamefont {Aiello}}, \bibinfo {author}
  {\bibfnamefont {M.~V.}\ \bibnamefont {Chekhova}}, \bibinfo {author}
  {\bibfnamefont {C.}~\bibnamefont {Silberhorn}}, \bibinfo {author}
  {\bibfnamefont {G.}~\bibnamefont {Leuchs}}, \ and\ \bibinfo {author}
  {\bibfnamefont {C.}~\bibnamefont {Marquardt}},\ }\href {\doibase
  10.1038/ncomms2838} {\bibfield  {journal} {\bibinfo  {journal} {Nature
  Communications}\ }\textbf {\bibinfo {volume} {4}},\ \bibinfo {pages} {1818}
  (\bibinfo {year} {2013})}\BibitemShut {NoStop}%
\bibitem [{\citenamefont {Liscidini}\ and\ \citenamefont
  {Andreani}(2004)}]{liscidini_highly_2004}%
  \BibitemOpen
  \bibfield  {author} {\bibinfo {author} {\bibfnamefont {M.}~\bibnamefont
  {Liscidini}}\ and\ \bibinfo {author} {\bibfnamefont {L.~C.}\ \bibnamefont
  {Andreani}},\ }\href {\doibase doi:10.1063/1.1786657} {\bibfield  {journal}
  {\bibinfo  {journal} {Applied Physics Letters}\ }\textbf {\bibinfo {volume}
  {85}},\ \bibinfo {pages} {1883} (\bibinfo {year} {2004})}\BibitemShut
  {NoStop}%
\bibitem [{\citenamefont {Zhang}\ \emph {et~al.}(2009)\citenamefont {Zhang},
  \citenamefont {{McCutcheon}}, \citenamefont {Burgess},\ and\ \citenamefont
  {Lon\v{c}ar}}]{zhang_ultra-high-Q_2009}%
  \BibitemOpen
  \bibfield  {author} {\bibinfo {author} {\bibfnamefont {Y.}~\bibnamefont
  {Zhang}}, \bibinfo {author} {\bibfnamefont {M.~W.}\ \bibnamefont
  {{McCutcheon}}}, \bibinfo {author} {\bibfnamefont {I.~B.}\ \bibnamefont
  {Burgess}}, \ and\ \bibinfo {author} {\bibfnamefont {M.}~\bibnamefont
  {Lon\v{c}ar}},\ }\href {\doibase 10.1364/OL.34.002694} {\bibfield  {journal}
  {\bibinfo  {journal} {Optics Letters}\ }\textbf {\bibinfo {volume} {34}},\
  \bibinfo {pages} {2694} (\bibinfo {year} {2009})}\BibitemShut {NoStop}%
\bibitem [{\citenamefont {Burgess}\ \emph {et~al.}(2009)\citenamefont
  {Burgess}, \citenamefont {Zhang}, \citenamefont {{McCutcheon}}, \citenamefont
  {Rodriguez}, \citenamefont {Bravo-Abad}, \citenamefont {Johnson},\ and\
  \citenamefont {Loncar}}]{burgess_design_2009}%
  \BibitemOpen
  \bibfield  {author} {\bibinfo {author} {\bibfnamefont {I.~B.}\ \bibnamefont
  {Burgess}}, \bibinfo {author} {\bibfnamefont {Y.}~\bibnamefont {Zhang}},
  \bibinfo {author} {\bibfnamefont {M.~W.}\ \bibnamefont {{McCutcheon}}},
  \bibinfo {author} {\bibfnamefont {A.~W.}\ \bibnamefont {Rodriguez}}, \bibinfo
  {author} {\bibfnamefont {J.}~\bibnamefont {Bravo-Abad}}, \bibinfo {author}
  {\bibfnamefont {S.~G.}\ \bibnamefont {Johnson}}, \ and\ \bibinfo {author}
  {\bibfnamefont {M.}~\bibnamefont {Loncar}},\ }\href {\doibase
  10.1364/OE.17.020099} {\bibfield  {journal} {\bibinfo  {journal} {Optics
  Express}\ }\textbf {\bibinfo {volume} {17}},\ \bibinfo {pages} {20099}
  (\bibinfo {year} {2009})}\BibitemShut {NoStop}%
\bibitem [{\citenamefont {Thon}\ \emph {et~al.}(2010)\citenamefont {Thon},
  \citenamefont {Irvine}, \citenamefont {Kleckner},\ and\ \citenamefont
  {Bouwmeester}}]{thon_polychromatic_2010}%
  \BibitemOpen
  \bibfield  {author} {\bibinfo {author} {\bibfnamefont {S.~M.}\ \bibnamefont
  {Thon}}, \bibinfo {author} {\bibfnamefont {W.~T.~M.}\ \bibnamefont {Irvine}},
  \bibinfo {author} {\bibfnamefont {D.}~\bibnamefont {Kleckner}}, \ and\
  \bibinfo {author} {\bibfnamefont {D.}~\bibnamefont {Bouwmeester}},\ }\href
  {\doibase 10.1103/PhysRevLett.104.243901} {\bibfield  {journal} {\bibinfo
  {journal} {Physical Review Letters}\ }\textbf {\bibinfo {volume} {104}},\
  \bibinfo {pages} {243901} (\bibinfo {year} {2010})}\BibitemShut {NoStop}%
\bibitem [{\citenamefont {Rivoire}, \citenamefont {Buckley},\ and\
  \citenamefont
  {Vu\v{c}kovi\'{c}}(2011{\natexlab{a}})}]{rivoire_multiply_2011}%
  \BibitemOpen
  \bibfield  {author} {\bibinfo {author} {\bibfnamefont {K.}~\bibnamefont
  {Rivoire}}, \bibinfo {author} {\bibfnamefont {S.}~\bibnamefont {Buckley}}, \
  and\ \bibinfo {author} {\bibfnamefont {J.}~\bibnamefont {Vu\v{c}kovi\'{c}}},\
  }\href {\doibase 10.1364/OE.19.022198} {\bibfield  {journal} {\bibinfo
  {journal} {Optics Express}\ }\textbf {\bibinfo {volume} {19}},\ \bibinfo
  {pages} {22198} (\bibinfo {year} {2011}{\natexlab{a}})}\BibitemShut {NoStop}%
\bibitem [{\citenamefont {Rivoire}, \citenamefont {Buckley},\ and\
  \citenamefont
  {Vu\v{c}kovi\'{c}}(2011{\natexlab{b}})}]{rivoire_multiply_2011-1}%
  \BibitemOpen
  \bibfield  {author} {\bibinfo {author} {\bibfnamefont {K.}~\bibnamefont
  {Rivoire}}, \bibinfo {author} {\bibfnamefont {S.}~\bibnamefont {Buckley}}, \
  and\ \bibinfo {author} {\bibfnamefont {J.}~\bibnamefont {Vu\v{c}kovi\'{c}}},\
  }\href {\doibase doi:10.1063/1.3607281} {\bibfield  {journal} {\bibinfo
  {journal} {Applied Physics Letters}\ }\textbf {\bibinfo {volume} {99}},\
  \bibinfo {pages} {013114} (\bibinfo {year} {2011}{\natexlab{b}})}\BibitemShut
  {NoStop}%
\bibitem [{\citenamefont {Joannopoulos}\ \emph {et~al.}(2011)\citenamefont
  {Joannopoulos}, \citenamefont {Johnson}, \citenamefont {Winn},\ and\
  \citenamefont {Meade}}]{joannopoulos_photonic_2011}%
  \BibitemOpen
  \bibfield  {author} {\bibinfo {author} {\bibfnamefont {J.~D.}\ \bibnamefont
  {Joannopoulos}}, \bibinfo {author} {\bibfnamefont {S.~G.}\ \bibnamefont
  {Johnson}}, \bibinfo {author} {\bibfnamefont {J.~N.}\ \bibnamefont {Winn}}, \
  and\ \bibinfo {author} {\bibfnamefont {R.~D.}\ \bibnamefont {Meade}},\
  }\href@noop {} {\emph {\bibinfo {title} {Photonic Crystals: Molding the Flow
  of Light (Second Edition)}}}\ (\bibinfo  {publisher} {Princeton University
  Press},\ \bibinfo {city} {Princeton},\ \bibinfo {year} {2011})\BibitemShut {NoStop}%
\bibitem [{\citenamefont {Takayama}\ \emph {et~al.}(2005)\citenamefont
  {Takayama}, \citenamefont {Kitagawa}, \citenamefont {Tanaka}, \citenamefont
  {Asano},\ and\ \citenamefont {Noda}}]{takayama_experimental_2005}%
  \BibitemOpen
  \bibfield  {author} {\bibinfo {author} {\bibfnamefont {S.-i.}\ \bibnamefont
  {Takayama}}, \bibinfo {author} {\bibfnamefont {H.}~\bibnamefont {Kitagawa}},
  \bibinfo {author} {\bibfnamefont {Y.}~\bibnamefont {Tanaka}}, \bibinfo
  {author} {\bibfnamefont {T.}~\bibnamefont {Asano}}, \ and\ \bibinfo {author}
  {\bibfnamefont {S.}~\bibnamefont {Noda}},\ }\href {\doibase
  doi:10.1063/1.2009060} {\bibfield  {journal} {\bibinfo  {journal} {Applied
  Physics Letters}\ }\textbf {\bibinfo {volume} {87}},\ \bibinfo {pages}
  {061107} (\bibinfo {year} {2005})}\BibitemShut {NoStop}%
\bibitem [{\citenamefont {{McCutcheon}}\ \emph {et~al.}(2011)\citenamefont
  {{McCutcheon}}, \citenamefont {Deotare}, \citenamefont {Zhang},\ and\
  \citenamefont {Lončar}}]{mccutcheon_high-q_2011}%
  \BibitemOpen
  \bibfield  {author} {\bibinfo {author} {\bibfnamefont {M.~W.}\ \bibnamefont
  {{McCutcheon}}}, \bibinfo {author} {\bibfnamefont {P.~B.}\ \bibnamefont
  {Deotare}}, \bibinfo {author} {\bibfnamefont {Y.}~\bibnamefont {Zhang}}, \
  and\ \bibinfo {author} {\bibfnamefont {M.}~\bibnamefont {Lončar}},\ }\href
  {\doibase doi:10.1063/1.3568897} {\bibfield  {journal} {\bibinfo  {journal}
  {Applied Physics Letters}\ }\textbf {\bibinfo {volume} {98}},\ \bibinfo
  {pages} {111117} (\bibinfo {year} {2011})}\BibitemShut {NoStop}%
\bibitem [{\citenamefont {Dvorak}\ \emph {et~al.}(1994)\citenamefont
    {Dvorak},
  \citenamefont {Schroeder}, \citenamefont {Andersen}, \citenamefont {Smirl},\
  and\ \citenamefont {Wherrett}}]{dvorak_measurement_1994}%
  \BibitemOpen
  \bibfield  {author} {\bibinfo {author} {\bibfnamefont {M.}~\bibnamefont
  {Dvorak}}, \bibinfo {author} {\bibfnamefont {W.}~\bibnamefont {Schroeder}},
  \bibinfo {author} {\bibfnamefont {D.}~\bibnamefont {Andersen}}, \bibinfo
  {author} {\bibfnamefont {A.~L.}\ \bibnamefont {Smirl}}, \ and\ \bibinfo
  {author} {\bibfnamefont {B.}~\bibnamefont {Wherrett}},\ }\href {\doibase
  10.1109/3.283768} {\bibfield  {journal} {\bibinfo  {journal} {{IEEE} Journal
  of Quantum Electronics}\ }\textbf {\bibinfo {volume} {30}},\ \bibinfo {pages}
  {256} (\bibinfo {year} {1994})}\BibitemShut {NoStop}%
\bibitem [{\citenamefont {Altug}\ and\ \citenamefont
  {Vuckovic}(2005)}]{altug_photonic_2005}%
  \BibitemOpen
  \bibfield  {author} {\bibinfo {author} {\bibfnamefont {H.}~\bibnamefont
  {Altug}}\ and\ \bibinfo {author} {\bibfnamefont {J.}~\bibnamefont
  {Vuckovic}},\ }\href {\doibase 10.1364/OPEX.13.008819} {\bibfield  {journal}
  {\bibinfo  {journal} {Optics Express}\ }\textbf {\bibinfo {volume} {13}},\
  \bibinfo {pages} {8819} (\bibinfo {year} {2005})}\BibitemShut {NoStop}%
\bibitem [{\citenamefont {Shoji}\ \emph {et~al.}(1997)\citenamefont {Shoji},
  \citenamefont {Kondo}, \citenamefont {Kitamoto}, \citenamefont {Shirane},\
  and\ \citenamefont {Ito}}]{shoji_absolute_1997}%
  \BibitemOpen
  \bibfield  {author} {\bibinfo {author} {\bibfnamefont {I.}~\bibnamefont
  {Shoji}}, \bibinfo {author} {\bibfnamefont {T.}~\bibnamefont {Kondo}},
  \bibinfo {author} {\bibfnamefont {A.}~\bibnamefont {Kitamoto}}, \bibinfo
  {author} {\bibfnamefont {M.}~\bibnamefont {Shirane}}, \ and\ \bibinfo
  {author} {\bibfnamefont {R.}~\bibnamefont {Ito}},\ }\href {\doibase
  10.1364/JOSAB.14.002268} {\bibfield  {journal} {\bibinfo  {journal} {Journal
  of the Optical Society of America B}\ }\textbf {\bibinfo {volume} {14}},\
  \bibinfo {pages} {2268} (\bibinfo {year} {1997})}\BibitemShut {NoStop}%
\bibitem [{buc()}]{buckley_suppinfo_2013}%
  \BibitemOpen
  \href@noop {} {}\bibinfo {note} {See supplementary material at [URL will be
  inserted by APL] for information on measurements of second harmonic
  generation in perturbed L3 photonic crystal cavities.}\BibitemShut {Stop}%
\bibitem [{\citenamefont {Bennett}, \citenamefont {Soref},\ and\
    \citenamefont
  {del Alamo}(1990)}]{bennett_carrier-induced_1990}%
  \BibitemOpen
  \bibfield  {author} {\bibinfo {author} {\bibfnamefont {B.}~\bibnamefont
  {Bennett}}, \bibinfo {author} {\bibfnamefont {R.~A.}\ \bibnamefont {Soref}},
  \ and\ \bibinfo {author} {\bibfnamefont {J.}~\bibnamefont {del Alamo}},\
  }\href {\doibase 10.1109/3.44924} {\bibfield  {journal} {\bibinfo  {journal}
  {{IEEE} Journal of Quantum Electronics}\ }\textbf {\bibinfo {volume} {26}},\
  \bibinfo {pages} {113} (\bibinfo {year} {1990})}\BibitemShut {NoStop}%
\bibitem [{\citenamefont {Gibbs}(1985)}]{gibbs_chapter_1985}%
  \BibitemOpen
  \bibfield  {author} {\bibinfo {author} {\bibfnamefont {H.~M.}\ \bibnamefont
  {Gibbs}},\ }in\ \href
  {http://www.sciencedirect.com/science/article/pii/B9780122819407500111}
  {\emph {\bibinfo {booktitle} {Optical Bistability: Controlling Light with
  Light}}}\ (\bibinfo  {publisher} {Academic Press},\ \bibinfo {year} {1985})\
  pp.\ \bibinfo {pages} {241--303}\BibitemShut {NoStop}%
\bibitem [{\citenamefont {Van}\ \emph {et~al.}(2002)\citenamefont {Van},
  \citenamefont {Ibrahim}, \citenamefont {Absil}, \citenamefont {Johnson},
  \citenamefont {Grover},\ and\ \citenamefont {Ho}}]{van_optical_2002}%
  \BibitemOpen
  \bibfield  {author} {\bibinfo {author} {\bibfnamefont {V.}~\bibnamefont
  {Van}}, \bibinfo {author} {\bibfnamefont {T.}~\bibnamefont {Ibrahim}},
  \bibinfo {author} {\bibfnamefont {P.}~\bibnamefont {Absil}}, \bibinfo
  {author} {\bibfnamefont {F.}~\bibnamefont {Johnson}}, \bibinfo {author}
  {\bibfnamefont {R.}~\bibnamefont {Grover}}, \ and\ \bibinfo {author}
  {\bibfnamefont {P.-T.}\ \bibnamefont {Ho}},\ }\href {\doibase
  10.1109/JSTQE.2002.1016376} {\bibfield  {journal} {\bibinfo  {journal}
  {{IEEE} Journal of Selected Topics in Quantum Electronics}\ }\textbf
  {\bibinfo {volume} {8}},\ \bibinfo {pages} {705 } (\bibinfo {year}
  {2002})}\BibitemShut {NoStop}%
\bibitem [{\citenamefont {Xu}\ and\ \citenamefont
  {Lipson}(2006)}]{xu_carrier-induced_2006}%
  \BibitemOpen
  \bibfield  {author} {\bibinfo {author} {\bibfnamefont {Q.}~\bibnamefont
  {Xu}}\ and\ \bibinfo {author} {\bibfnamefont {M.}~\bibnamefont {Lipson}},\
  }\href {\doibase 10.1364/OL.31.000341} {\bibfield  {journal} {\bibinfo
  {journal} {Optics Letters}\ }\textbf {\bibinfo {volume} {31}},\ \bibinfo
  {pages} {341} (\bibinfo {year} {2006})}\BibitemShut {NoStop}%
\bibitem [{\citenamefont {Almeida}\ \emph {et~al.}(2004)\citenamefont
  {Almeida}, \citenamefont {Barrios}, \citenamefont {Panepucci},\ and\
  \citenamefont {Lipson}}]{almeida_all-optical_2004}%
  \BibitemOpen
  \bibfield  {author} {\bibinfo {author} {\bibfnamefont {V.~R.}\ \bibnamefont
  {Almeida}}, \bibinfo {author} {\bibfnamefont {C.~A.}\ \bibnamefont
  {Barrios}}, \bibinfo {author} {\bibfnamefont {R.~R.}\ \bibnamefont
  {Panepucci}}, \ and\ \bibinfo {author} {\bibfnamefont {M.}~\bibnamefont
  {Lipson}},\ }\href {\doibase 10.1038/nature02921} {\bibfield  {journal}
  {\bibinfo  {journal} {Nature}\ }\textbf {\bibinfo {volume} {431}},\ \bibinfo
  {pages} {1081} (\bibinfo {year} {2004})}\BibitemShut {NoStop}%
\bibitem [{\citenamefont {Nozaki}\ \emph {et~al.}(2010)\citenamefont
    {Nozaki},
  \citenamefont {Tanabe}, \citenamefont {Shinya}, \citenamefont {Matsuo},
  \citenamefont {Sato}, \citenamefont {Taniyama},\ and\ \citenamefont
  {Notomi}}]{nozaki_sub-femtojoule_2010}%
  \BibitemOpen
  \bibfield  {author} {\bibinfo {author} {\bibfnamefont {K.}~\bibnamefont
  {Nozaki}}, \bibinfo {author} {\bibfnamefont {T.}~\bibnamefont {Tanabe}},
  \bibinfo {author} {\bibfnamefont {A.}~\bibnamefont {Shinya}}, \bibinfo
  {author} {\bibfnamefont {S.}~\bibnamefont {Matsuo}}, \bibinfo {author}
  {\bibfnamefont {T.}~\bibnamefont {Sato}}, \bibinfo {author} {\bibfnamefont
  {H.}~\bibnamefont {Taniyama}}, \ and\ \bibinfo {author} {\bibfnamefont
  {M.}~\bibnamefont {Notomi}},\ }\href {\doibase 10.1038/nphoton.2010.89}
  {\bibfield  {journal} {\bibinfo  {journal} {Nature Photonics}\ }\textbf
  {\bibinfo {volume} {4}},\ \bibinfo {pages} {477} (\bibinfo {year}
  {2010})}\BibitemShut {NoStop}%
\bibitem [{\citenamefont {Fan}\ \emph {et~al.}(2012)\citenamefont {Fan},
  \citenamefont {Wang}, \citenamefont {Varghese}, \citenamefont {Shen},
  \citenamefont {Niu}, \citenamefont {Xuan}, \citenamefont {Weiner},\ and\
  \citenamefont {Qi}}]{fan_all-silicon_2012}%
  \BibitemOpen
  \bibfield  {author} {\bibinfo {author} {\bibfnamefont {L.}~\bibnamefont
  {Fan}}, \bibinfo {author} {\bibfnamefont {J.}~\bibnamefont {Wang}}, \bibinfo
  {author} {\bibfnamefont {L.~T.}\ \bibnamefont {Varghese}}, \bibinfo {author}
  {\bibfnamefont {H.}~\bibnamefont {Shen}}, \bibinfo {author} {\bibfnamefont
  {B.}~\bibnamefont {Niu}}, \bibinfo {author} {\bibfnamefont {Y.}~\bibnamefont
  {Xuan}}, \bibinfo {author} {\bibfnamefont {A.~M.}\ \bibnamefont {Weiner}}, \
  and\ \bibinfo {author} {\bibfnamefont {M.}~\bibnamefont {Qi}},\ }\href
  {\doibase 10.1126/science.1214383} {\bibfield  {journal} {\bibinfo  {journal}
  {Science}\ }\textbf {\bibinfo {volume} {335}},\ \bibinfo {pages} {447}
  (\bibinfo {year} {2012})},\ \bibinfo {note} {{PMID:} 22194410}\BibitemShut
  {NoStop}%
\bibitem [{\citenamefont {Lee}\ \emph {et~al.}(2009)\citenamefont {Lee},
  \citenamefont {Kiravittaya}, \citenamefont {Kumar}, \citenamefont {Plumhof},
  \citenamefont {Balet}, \citenamefont {Li}, \citenamefont {Francardi},
  \citenamefont {Gerardino}, \citenamefont {Fiore}, \citenamefont {Rastelli},\
  and\ \citenamefont {Schmidt}}]{lee_local_2009}%
  \BibitemOpen
  \bibfield  {author} {\bibinfo {author} {\bibfnamefont {H.~S.}\ \bibnamefont
  {Lee}}, \bibinfo {author} {\bibfnamefont {S.}~\bibnamefont {Kiravittaya}},
  \bibinfo {author} {\bibfnamefont {S.}~\bibnamefont {Kumar}}, \bibinfo
  {author} {\bibfnamefont {J.~D.}\ \bibnamefont {Plumhof}}, \bibinfo {author}
  {\bibfnamefont {L.}~\bibnamefont {Balet}}, \bibinfo {author} {\bibfnamefont
  {L.~H.}\ \bibnamefont {Li}}, \bibinfo {author} {\bibfnamefont
  {M.}~\bibnamefont {Francardi}}, \bibinfo {author} {\bibfnamefont
  {A.}~\bibnamefont {Gerardino}}, \bibinfo {author} {\bibfnamefont
  {A.}~\bibnamefont {Fiore}}, \bibinfo {author} {\bibfnamefont
  {A.}~\bibnamefont {Rastelli}}, \ and\ \bibinfo {author} {\bibfnamefont
  {O.~G.}\ \bibnamefont {Schmidt}},\ }\href {\doibase doi:10.1063/1.3262961}
  {\bibfield  {journal} {\bibinfo  {journal} {Applied Physics Letters}\
  }\textbf {\bibinfo {volume} {95}},\ \bibinfo {pages} {191109} (\bibinfo
  {year} {2009})}\BibitemShut {NoStop}%

\end{thebibliography}

\end{document}